# Long-lived electron spin coherence in Ga-doped crystals at room temperature

Zhen Wu,[1] Meizhen Jiang,[1] Jiaxing Guo,[1] Qing Yang,[1] Yuanyuan Zhang,[1] Tianqing Jia,[1] Zhenrong Sun,[1] and Donghai Feng[1,2,*]

[1]*State Key Laboratory of Precision Spectroscopy, East China Normal University, Shanghai 200241, China*

[2]*Collaborative Innovation Center of Extreme Optics, Shanxi University, Taiyuan, Shanxi 030006, China*

Abstract: Electron spin dynamics are studied in Ga-doped ZnO single crystals by time-resolved Faraday and Kerr rotation spectroscopies. Long-lived spin coherence with two dephasing processes is discovered where the characteristic time is up to 5.2 ns at room temperature. Through the dependence measurements of laser wavelength and temperature, the room-temperature long-lived spin signal is attributed to localized electrons. The spin dephasing (relaxation) processes are independent of transverse (longitudinal) magnetic fields, indicating the spin dephasing not resulting from the *g*-factor inhomogeneity and electron-nuclear hyperfine interaction. It reveals that the two spin dephasing processes originate from two types of localized electrons, both of which are dominated by the anisotropic exchange Dzyaloshinskii-Moriya interaction between adjacent localized electrons.

[*]dhfeng@phy.ecnu.edu.cn

Spintronics and spin-based quantum information processing are the exploration of electron spin degree of freedom to realize ultra-high speed information processing and quantum computation [1,2]. Long-lived spin coherence is required in information manipulation and storage in practical applications, ideally if it can be achieved in conventional semiconductors at room temperature. ZnO is a wide and direct bandgap semiconductor with weak spin-orbit coupling and weak electron-nuclear hyperfine interaction. It is expected to achieve a long electron spin coherence time. The electron spin coherence dynamics have been experimentally studied in various ZnO materials from bulk crystal to quantum dots [3–7]. The electron spin dephasing time of 25 ns has been discovered by electron paramagnetic resonance spectroscopy in ZnO quantum dots at room temperature [4]. Whereas, optically-detected electron spin coherence time at room temperature is only up to ~190 ps in bulk ZnO [3] and ~1.2 ns in ZnO sol-gel films [5]. To date, no longer room-temperature electron spin coherence has been reported in ZnO-based materials by time-resolved optical spectroscopies.

In this work, we use time-resolved Faraday rotation (TRFR) and Kerr rotation (TRKR) spectroscopies to study the electron spin coherence dynamics in Ga-doped ZnO single crystals, and find that the electron spin dephasing time is up to 5.2 ns at room temperature. The electron spin coherence in Ga-doped ZnO bulk crystals is quite robust with temperature and magnetic fields. We exclude the possibility of the spin origin from itinerant electrons and attribute it to localized electrons by laser wavelength and temperature dependence measurements. Localized electrons in Ga-doped ZnO bulk crystals show advantages in room-temperature spin coherence as compared with other traditional III-V and II-VI bulk semiconductors, both doped and undoped [8–16], where shorter spin coherence time has been reported at room temperature (e.g., ~110 ps in bulk GaAs [8], ~100 ps in bulk wurtzite GaN [10], ~2.5 ns in bulk cubic GaN [11], and ~60 ps in bulk CdTe [12]) or long-lived spin coherence has been typically reported at low temperature.

The sample used in the experiments is an n-type Ga-doped ZnO single crystal in the wurtzite structure prepared by hydrothermal methods and commercially obtained from Hefei Yuanjing Technology Materials Co., Ltd. The crystal surface is (0001) with a thickness of 1 mm and double side polished. The electron concentration is $1.58\times10^{18}$ cm$^{-3}$. The electron spin dynamics are investigated by TRFR/TRKR spectroscopies, which are often used to measure electron spin dynamics in various semiconductors [17–19]. The laser source is a femtosecond laser amplifier (Pharos, Light Conversion) with a central wavelength of 1030 nm and a repetition frequency of 50 kHz. The output of Pharos is split into two beams. One beam passes through a femtosecond optical parametric amplifier (Orpheus) and a frequency doubling crystal BBO as the pump light with a spectral width of ~109 cm$^{-1}$ and a pulse width of ~190 fs. The other beam passes through a second harmonic bandwidth compressor and a picosecond optical parametric amplifier (Orpheus-PS) as the probe light with a spectral width of ~9.9 cm$^{-1}$ and a pulse width of ~3.1 ps. The pump pulses are modulated between $\sigma^+$ and $\sigma^-$ circular polarizations at a frequency of 20 kHz by an electro-optical modulator and generate transient electron spin polarization in the sample. The subsequent electron spin dynamics are monitored by the rotation angle of the polarization plane of the linearly polarized probe pulses in either transmission (TRFR) or reflection mode (TRKR). The FR and KR signals are recorded using an optical polarization bridge in conjunction with lock-in detection. The time delays between pump and probe pulses are adjusted by a mechanical delay line. The temperature dependence measurements are performed in a closed cycle optical cryostat, with the temperature adjustable from 5 K to 300 K. The photoluminescence (PL) is excited by a 360 nm CW laser, and collected by an EMCCD spectrometer.

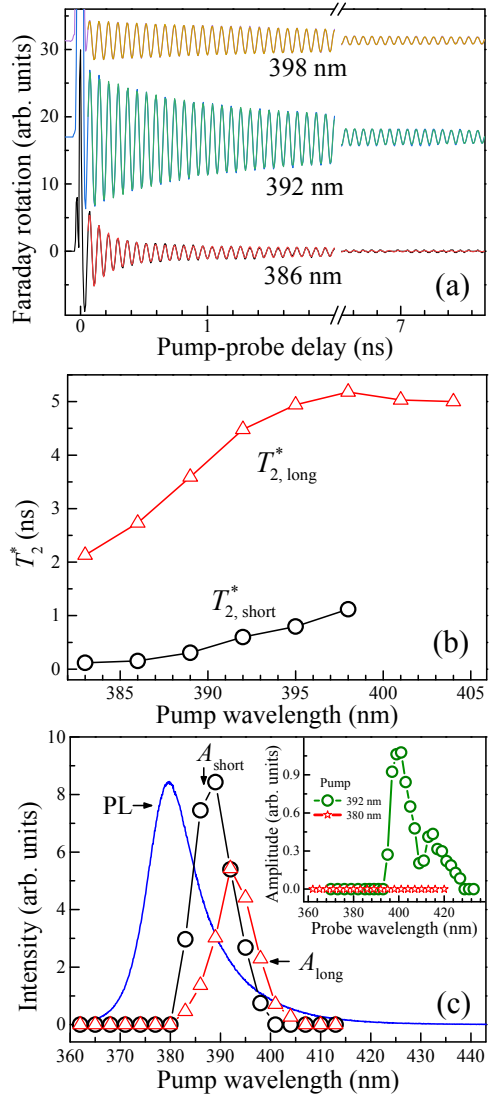

FIG. 1. (a) TRFR signals at different pump wavelengths with the probe wavelength fixed at 401 nm at room temperature. The spin signals can be well fitted by the function $\theta(t) = \cos(2\pi\nu_L t)\left[A_{\text{short}}\exp\left(-\frac{t}{T^*_{2,\text{short}}}\right) + A_{\text{long}}\exp\left(-\frac{t}{T^*_{2,\text{long}}}\right)\right]$. The spin dynamics are continuously measured up to 7.7 ns and a break from 2 ns to 6.5 ns is made in the curves for better clarity of the periodic oscillations. (b) Spin dephasing time as a function of pump wavelength. (c) Spin amplitude for $A_{\text{short}}$ (black circle) and $A_{\text{long}}$ (red triangle) as a function of pump wavelength and PL spectrum (blue line). The inset shows the dependence of total spin amplitude ($A_{\text{short}} + A_{\text{long}}$) on probe wavelengths measured by TRKR spectroscopies with the pump wavelength fixed at 392 nm (olive circle) and 380 nm (red star), respectively.

Figure 1(a) shows the TRFR spectroscopy in Ga-doped ZnO single crystals in a transverse magnetic field of 500 mT at room temperature for different pump wavelengths with a fixed probe wavelength at 401 nm. The time-dependent rotation signals can be well fitted by the function

$$\theta(t) = \cos(2\pi\nu_L t)\left[A_{\text{short}}\exp\left(-\frac{t}{T^*_{2,\text{short}}}\right) + A_{\text{long}}\exp\left(-\frac{t}{T^*_{2,\text{long}}}\right)\right], \qquad (1)$$

where $\nu_L$ is the Larmor precession frequency. $T^*_{2,\text{short}}$ and $T^*_{2,\text{long}}$ are the spin dephasing times corresponding to the fast process with an initial amplitude of $A_{\text{short}}$ and the slow process with an initial amplitude of $A_{\text{long}}$, respectively. As shown in Fig. 1(b), both spin dephasing times increase with increasing pump wavelength, where $T^*_{2,\text{long}}$ reaches 5.2 ns at the pump wavelength of 398 nm and then levels off, and $T^*_{2,\text{short}}$ reaches 1.1 ns at 398 nm pump wavelength. Note that the envelopes of the spin signals at 401 nm and 404 nm are well fitted by a single exponential function with negligible fast dephasing processes. As shown in Fig. 1(c), the amplitudes of the fast and slow processes have a peak at the pump wavelength of 389 nm and 392 nm, respectively, both showing a considerable redshift to the PL peak of 380 nm. It indicates that the observed spin coherence signals in Fig. 1 are attributed to localized electrons. Below the pump wavelength of 380 nm, there are no detectable spin signals. The inset in Fig. 1(c) shows the total spin amplitude as a function of probe wavelength with the pump wavelength fixed at 392 nm and 380 nm, respectively, which is measured in reflection mode by TRKR spectroscopies (TRFR is not used because there is no transmission for photon energies above the bandgap). With the 392 nm pump, the spin amplitude shows a maximum at the probe wavelength of 401 nm. When the pump wavelength is 380 nm, i.e., PL peak wavelength, there are no spin signals detected.

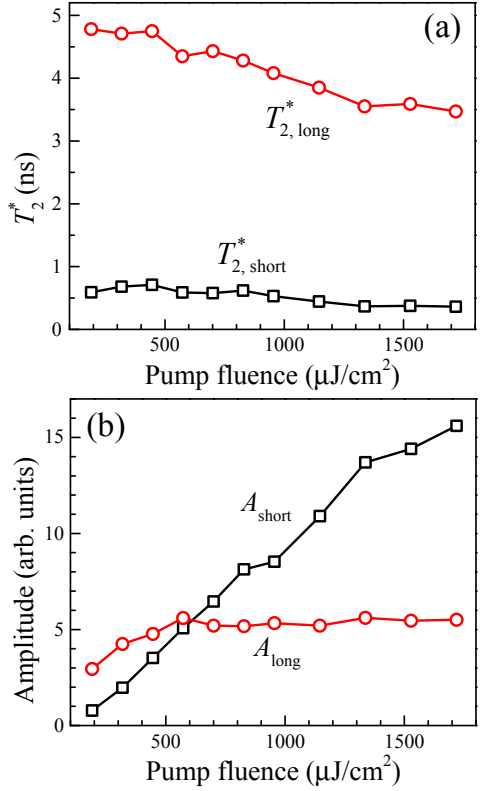

FIG. 2. (a) Spin dephasing time and (b) spin amplitude as a function of the pump fluence in the TRFR measurements. The pump and probe wavelengths are 392 nm and 401 nm, respectively. $B = 500$ mT and $T = 295$ K.

Figure 2 shows the pump fluence dependence of the spin dephasing time and amplitude in the TRFR measurements. The pump and probe wavelengths are set at 392 nm and 401 nm, respectively. We use TRFR here rather than TRKR measurements due to the fact that TRFR has a signal-to-noise ratio ~5 times stronger than TRKR for the 401 nm probe. As shown in Fig. 2(a), both the two spin dephasing times decrease with the increasing pump fluences. The amplitudes corresponding to the two dephasing processes show different trends. $A_{\text{long}}$ increases with the increasing fluences at first and then saturates above ~ 600 μJ/cm$^2$, while $A_{\text{short}}$ increases linearly with the increasing fluences and no obvious saturation is observed, as shown in Fig. 2(b). Further discussions will be made below in combination with the analysis of the spin dephasing mechanisms.

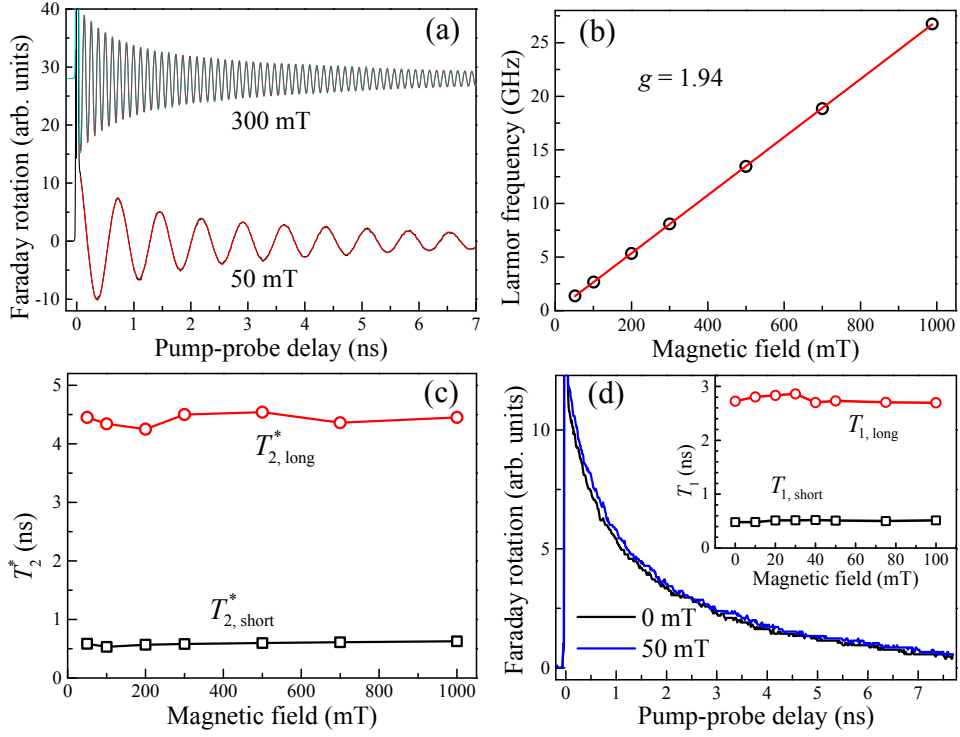

FIG. 3. (a) TRFR signals in different transverse magnetic fields at room temperature. The pump and probe wavelengths are 392 nm and 401 nm, respectively. (b) Larmor precession frequency as a function of transverse magnetic field. (c) Dependence of spin dephasing time $T_2^*$ on transverse magnetic fields. (d) TRFR signals in longitudinal magnetic fields of 0 mT and 50 mT. Inset shows the dependence of spin relaxation time $T_1$ on longitudinal magnetic fields.

Figure 3(a) shows the spin coherence dynamics in different transverse magnetic fields. All the curves can be well fitted by Eq. (1). The magnetic field dependence of Larmor precession frequencies is described by the equation $h\nu_L = \mu_B g B$, where $g$, $h$ and $\mu_B$ are the electron $g$ factor, the Planck constant and the Bohr magneton, respectively. $g = 1.94$ is evaluated from the linear fit in Fig. 3(b). Both dephasing times are independent of transverse magnetic fields, as shown in Fig. 3(c). It means that *g*-factor inhomogeneity is weak and inhomogeneous dephasing mechanism can be excluded in this sample. Figure 3(d) shows the longitudinal magnetic field dependence of spin dynamics. The dynamic curves are almost identical between 0 and 50 mT. The spin relaxation times $T_{1,\text{short}}$ and $T_{1,\text{long}}$ are independent of longitudinal magnetic field as shown in the inset of Fig. 3(d). Therefore, the hyperfine-induced spin relaxation can be excluded as it can be strongly suppressed by a small longitudinal magnetic field [20]. Interestingly, as shown in Figs. 3(c) and the inset of 3(d), $T_2^*$ of both fast and slow dephasing processes ($T_2^* = T_2$ due to the lack of inhomogeneous dephasing) is longer than $T_1$, and especially $T_{2,\text{long}}^* \approx 1.6\, T_{1,\text{long}}$, close to the $T_2 = 2T_1$ limit.

After exclusion of the *g*-factor inhomogeneity and the electron-nuclear hyperfine interaction, the spin dephasing mechanism can be only the anisotropic exchange Dzyaloshinskii-Moriya (DM) interaction between adjacent localized electrons [21-24]. The DM interaction is magnetic field independent. There are two types of localized electrons ($e_1$ and $e_2$) in Ga-doped ZnO single crystals, responsible for the two dephasing/relaxation processes. $e_1$ has a binding energy around 75 meV corresponding to the spin peak of 389 nm, and $e_2$ has a binding energy of around 100 meV corresponding to the spin peak of 392 nm [Fig. 1(c)]. We speculate that $e_1$ has a higher concentration than $e_2$, leading to a faster spin dephasing. It is supported by the fact that the spin signals of $e_1$ both at the peak wavelength [Fig. 1(c)] and at high pump fluences [Fig. 2(b)] are stronger than those of $e_2$. Note that the spin signals of $e_1$ at low pump fluences are weaker than those of $e_2$ as shown in Fig. 2(b), due to the fact that the pump wavelength of 392 nm is resonant to the $e_2$ excitation but off-resonant to the $e_1$ excitation.

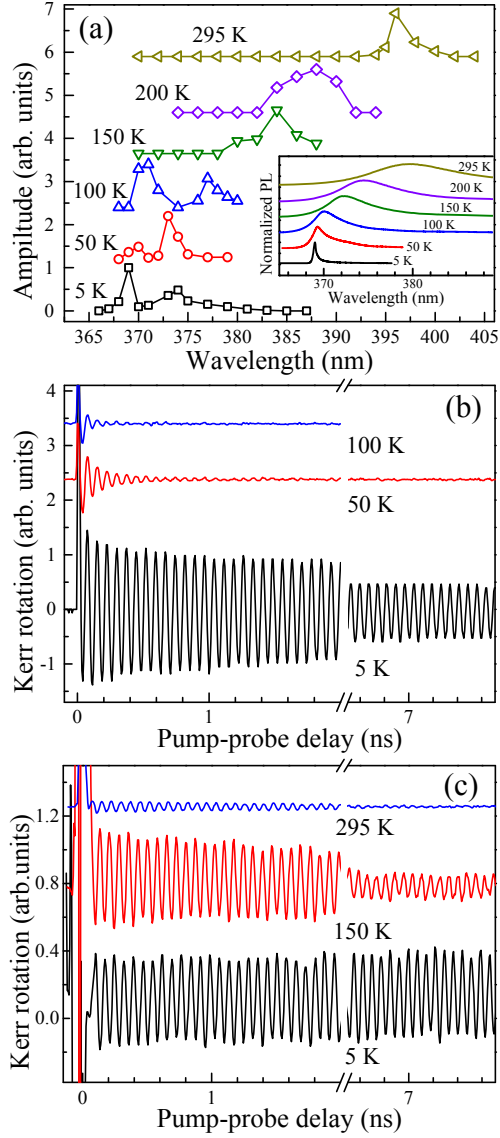

FIG. 4. Temperature dependence of TRKR signals at degenerate pump/probe wavelengths in a transverse magnetic field of 500 mT. (a) Wavelength dependence of total spin amplitude at different temperatures. The amplitude is normalized by the maximum. Inset shows the normalized PL spectra at different temperatures. (b) Electron spin dynamics of itinerant electrons and (c) of localized electrons measured at the peak wavelengths at different temperatures. The spin dynamics in (b) and (c) are continuously measured up to 7.7 ns and a break from 2 ns to 6.5 ns is made in the curves for better clarity of the periodic oscillations.

Figure 4(a) shows the wavelength dependence of spin amplitudes at different temperature. The measurements are performed with degenerate pump/probe wavelength by TRKR spectroscopies. At low temperature (<100 K), there are two peaks. The first peak of the spin amplitude is in line with the PL peak, which is attributed to itinerant electrons in the conduction band. Above 100 K, the first peak disappears and no spin signals of itinerant electrons are observed. The second spin peak is redshift to the first spin peak (or PL peak) and attributed to localized electrons. With increasing the temperature, both spin peaks have a redshift, which is caused by the bandgap narrowing. Compared with the first peak, the second peak has a stronger redshift. Figures 4(b) and 4(c) show the spin coherence dynamics of itinerant and localized electrons at different temperatures, respectively. At 5 K, both itinerant and localized electrons have long-lived spin coherence. The spin dephasing time is ~9.2 ns for itinerant electrons and too long to be evaluated for localized electrons, because the spin of localized electrons has no obvious decay within the measurement range. Both the spin amplitude and dephasing time of the itinerant electrons decrease with the increasing temperature. The spin dephasing of itinerant electrons is dominated by D'yakonov-Perel' (DP) mechanism [23,25]. When the temperature is above 100 K, the spin of itinerant electrons relaxes fast and cannot be detected any more. In contrast, the spin coherence of localized electrons is more robust, and detectable even at room temperature.

In conclusion, we have discovered long-lived spin coherence of localized electrons in Ga-doped ZnO single crystals at room temperature. The electron spin dephasing has two characteristic times which are independent of transverse magnetic fields, indicating the spin dephasing is not dominated by *g*-factor inhomogeneities. The dependence measurements of longitudinal magnetic fields exclude the dephasing and relaxation mechanism of electron-nuclear hyperfine interaction. We conclude that the two spin dephasing processes originate from two types of localized electrons. The anisotropic exchange DM interactions between adjacent localized electrons dominate both the electron spin dephasing processes.

This work is supported by the National Key Research and Development Program of China (Grants No. 2018YFA0306303), the National Natural Science Foundation of China (Grants No. 91950112, No. 11727810, No. 12074123, and No. 61720106009), and the Science and Technology Commission of Shanghai Municipality (Grants No. 19ZR1414500).